**Mapping and Comparing Climate Equity Policy Practices Using RAG LLM-Based Semantic Analysis and Recommendation Systems**

Seung Jun Choi[a]

[a] Department of Geography, Planning, and Recreation, College of Social and Behavioral Sciences, Northern Arizona University, Flagstaff, AZ, 86001, United States

**Abstract**

This study investigates the use of large language models to enhance the policymaking process. We first analyze planning-related job postings to revisit the evolving roles of planners in the era of AI. We then examine climate equity plans across the U.S. and apply ChatGPT to conduct semantic analysis, extracting policy, strategy, and action items related to transportation and energy. The methodological framework relied on a LangChain-native retrieval-augmented generation pipeline. Based on these extracted elements and their evaluated presence, we develop a content-based recommendation system to support cross-city policy comparison. The results indicate that, despite growing attention to AI, planning jobs largely retain their traditional domain emphases in transportation, environmental planning, housing, and land use. Communicative responsibilities remain central to planning practice. Climate equity plans commonly address transportation, environmental, and energy-related measures aimed at reducing greenhouse gas emissions and predominantly employ affirmative language. The demonstration of the recommendation system illustrates how planners can efficiently identify cities with similar policy practices, revealing patterns of geographic similarity in policy adoption. The study concludes by envisioning localized yet personalized AI-assisted systems that can be adapted within urban systems.

# 1. Introduction

Since the emergence of Large Language Models (LLMs), including ChatGPT, Gemini, and Claude, and the subsequent development of agentic tools, recent years have seen growing efforts to deploy these systems to partially or fully substitute for human labor. In their simplest form, LLMs have been integrated into chatbot interfaces to support information retrieval and response generation (Ramjee et al., 2025). More recently, they have been synchronized with physical infrastructure, including autonomous vehicles (Chiu et al., 2025), traffic signals (Lai et al., 2025), and smart city systems (Kalyuzhnaya et al., 2025). Human labor associated with repetitive tasks, particularly work traditionally assigned to junior workers, is increasingly being replaced by AI-driven systems (Brynjolfsson et al., 2025), indicating that private firms have actively adopted these technologies in business operations.

Against the backdrop of ongoing paradigm shifts and increasing uncertainty in the field, this study emerged from a fundamental question: can urban policy documentation and management be partially or fully performed by AI? More broadly, the study asks whether the work of planners and policymakers can be comprehensively evaluated, assessed, and supported through AI-driven systems. Planning is an inherently interdisciplinary field and, epistemologically, has long advocated for the needs of disadvantaged groups through traditions such as advocacy planning and equity planning (Brooks, 2002). This orientation imposes an obligation on planners to advance the public good, typically operationalized through plans and policies. While planners must engage directly with communities and participate in field-based, natural language interactions (Sager, 2017), and while AI is likely to significantly reshape the planning workforce, it is unlikely to fully replace the core functions of planners.

Nevertheless, what if repetitive tasks could be replaced by AI, allowing human planners to increase productivity? Planners are often actively engaged in discussions that shape policy implications, yet they can only partially review planning documents because it is impractical to examine all relevant policies worldwide. It is also difficult to objectively determine whether a policy implemented in one city has been adopted, adapted, or omitted in another. The present study was developed in response to these challenges and seeks to utilize AI, specifically LLMs, for policy review. Related endeavors encompass the utilization of these methodologies for legal document classification (Wei et al., 2023) and the development of conceptual frameworks advocating their application in scenario planning (Kgomo and Song, 2025). Recent research, such as the work by Deng et al. (2025), has explored the use of ChatGPT to automate the collection of local extreme heat policies. However, their study was limited in geographic scope to Miami and Chicago, reviewed only eight documents, and did not employ accuracy-enhancing methods such as Retrieval-Augmented Generation (RAG).

This study explores how LLMs can be utilized for public policy analysis, with a particular focus on evaluating the presence and distribution of existing policies across multiple cities. In countries with large and complex governance structures, such as the U.S. and China, policy authority is distributed across multiple hierarchical levels, creating substantial challenges for systematic policy comparison. In the U.S., the interaction between home rule and federal and state authority results in layered planning documents, including comprehensive plans, regional plans, and strategic plans. Given the central role of equity planning in contemporary planning practice (Brooks, 2002) and the growing emphasis on climate action to address climate change and advance sustainability goals (Aboagye and Sharifi, 2024; Franco et al., 2025), this study focuses on climate equity/justice and action plans in the U.S. In addition, the study revisits the professional obligations of planners by examining how planning work is evolving in response to AI adoption, drawing on a web-scraping analysis of planning-related job postings across the U.S.

The objective of this study is to demonstrate how similar policy practices can be automatically matched and policy gaps identified through a recommendation system. This study focuses on how planning practice can be reformed, rather than on evaluating which AI model performs better. We envision a system in which public agencies, government bureaus, or individual practitioners can automatically collect policy documents and compare them based on similarity metrics, analogous to recommendation mechanisms used by platforms such as Netflix, YouTube, or search engines. The remainder of this article is structured as follows. Following the introduction, Section 2 reviews the relevant literature and provides the rationale for focusing on equity planning and the use of AI in planning practice. Section 3 describes the materials and methods. Section 4 presents the analytical results. Section 5 discusses the findings and their implications. Finally, Section 6 concludes the paper with a summary of the main contributions and directions for future research.

## 2. Literature Review

### 2.1. Obligations of Planners: Equity and Climate Action Planning

The origins of equity planning date back to the 1970s, when Norman Krumholz advanced the idea of redistributing resources and emphasized the role of planners as institutional activists. Since then, equity planning has become a major normative paradigm and a core professional obligation for planners (Brooks, 2002). Equity planning is not static; rather, it has continually evolved to address the dominant social and political challenges of each era. From a retrospective perspective, Zapata and Bates (2015) argue that equity planning has expanded across multiple fields, including health equity, climate planning, affordable housing, and planning pedagogy. They emphasize that equity planning is not confined to formal planning bureaus. What matters most is not one's title, but the ability to build a collective vision and advance change across diverse institutional contexts, including community-based organizations, metropolitan planning organizations, and community activists (Zapata and Bates, 2015).

One of the urgent issues in contemporary planning is dealing with extreme climates. Climate extremes refer to infrequent weather or climate events identified relative to local historical climate conditions (Seneviratne et al., 2021). Ongoing climate change alters their likelihood and severity, increasing the occurrence of events that were previously rare or absent in the observational record (Seneviratne et al., 2021). Frequent exposure to natural hazards or irregular climate conditions is becoming the new normal. According to UN Environment (2022), recent climate observations indicate that temperature thresholds once considered exceptional are being exceeded with increasing regularity, reflecting a rapidly intensifying climate crisis.

Trenberth et al. (2015) argued that climate extremes should be assessed through their impacts, emphasizing how climate change amplifies damages given an event rather than focusing solely on causal attribution. This perspective introduces two complementary viewpoints. The first centers on environmental justice, emphasizing the need to address disproportionate impacts across communities. From this perspective, plan-making should be holistic, carefully considering both short- and long-term policy impacts while actively working to mitigate potential harms (Resnik, 2022). Meaningful participation by diverse community stakeholders and grassroots activists is essential, alongside recognition of procedural justice in decision-making processes (Schlosberg and Collins, 2014).

The second viewpoint emphasizes the necessity of structural change through concrete action. Karner et al. (2020) distinguished equity from justice, placing greater emphasis on the latter. Justice, in this framing, requires addressing fundamental structural barriers and unequal access to opportunities and is more closely aligned with bottom-up approaches to change (Karner et al., 2020). Within this context, climate action plans have emerged as policy instruments that translate climate concerns into actionable strategies for both mitigation, largely focused on reducing greenhouse gas emissions, and adaptation to evolving climate conditions (Stone et al., 2012).

Globally, climate action is closely connected to a wide range of development objectives, with implications extending across 16 of the UN's Sustainable Development Goals (Fuso Nerini et al., 2019). This body of evidence suggests that climate action should be conceptualized not merely as an environmental agenda, but as an integrated development strategy that is structurally linked to poverty reduction, public health, urban sustainability, and institutional capacity building (Fuso Nerini et al., 2019). Comparative research further indicates that local climate action planning tends to concentrate on a limited number of core sectors. Across diverse national contexts, transportation and energy systems consistently emerge as the central domains of mitigation efforts, while other policy areas play more complementary roles. This sectoral pattern is evident in local planning frameworks in the U.S. (Deetjen et al., 2018) and Denmark (Damsø et al., 2016), as well as in national-level climate governance in China (Wu, 2023), where decarbonization strategies are largely oriented toward energy system transformation and transport-related emissions.

## 2.2. The Use of AI in Normative Planning

Planning can broadly be categorized into positive and normative approaches. Normative planning engages in discussions of what planning practice *should* be (Brooks, 2002), drawing on multiple theoretical lenses to define what is considered equitable, justifiable, or sustainable. Equity planning represents one of the most prominent traditions within normative planning practice. Empirical analyses of equity concerns inherently involve multiple analytical frameworks. Such analyses often focus on identifying disproportionate burdens, including limited access to opportunities or heightened exposure to harms, experienced by specific socio-demographic groups (Karner and Niemeier, 2013). In transportation planning, equity analysis has largely centered on measuring accessibility inequality or accessibility poverty, approaches that are grounded in Rawlsian and sufficientarian philosophical traditions (Karner et al., 2025). These approaches are strongly influenced by earlier work in environmental planning, where scholars examined relative exposure to environmental harms, frequently operationalized through proximity-based measures (Bowen, 2002). Within this tradition, comparison groups are typically defined using frameworks derived from welfare and choice theory, spatial location theory, and empirical positivism (Bowen, 2002).

Normative planning practices are not rendered irrelevant in the era of AI; rather, they are being reconfigured and expanded through AI-enabled analytical and participatory capacities. Choi and Jiao (2024) argue that equity planning can evolve alongside AI by augmenting analytical practices while opening new pathways for community engagement. Their work demonstrates how AI-based transportation accessibility classifications can be integrated into interactive dashboards, enabling multiple stakeholders, including community members, to identify and interpret the spatial characteristics of their own locations. Beyond accessibility analysis, Generative AI (GenAI), including LLMs, has been shown to enhance educational literacy (James and Andrews, 2024), identify overlooked locations for site planning (Choi, 2025), and support further convergence in land use planning (Liu et al., 2025). Collectively, these applications align with Suh et al.'s (2023) argument that AI enables new forms of Human–AI (HAI) co-creation, fostering collaborative and iterative planning practices rather than replacing human judgment. At the same time, AI is already reshaping AI governance regimes in ways that diverge fundamentally from the smart city paradigm (Caprotti et al., 2024). In this context, the key challenge for planners is not whether to adopt AI, but how to define thresholds and conditions for AI adoption that responsibly reconfigure contemporary planning practice while preserving its normative commitments.

Additionally, recommendation systems are widely used in everyday contexts to support decision-making, such as purchasing products or selecting content on streaming platforms. As such, they constitute a critical component of personalized systems. Traditionally, platforms such as YouTube have relied on retrieval-and-ranking frameworks based on users' interaction histories, including viewing and liking behaviors, thereby establishing the foundational architecture for later deep learning–based recommendation systems (Davidson et al., 2010). More recent developments have extended these systems by embedding LLMs, enabling capabilities such as automated text description generation within recommendation engines (Acharya et al., 2023) and the simulation of user engagement patterns (Zhang et al., 2025). While recommendation systems have been extensively studied in commercial and media platforms, their application within the planning domain remains relatively underexplored.

Nevertheless, as urban systems become increasingly intelligent and data-driven, they are also likely to become more personalized, supporting both plan-making processes and everyday urban experiences. Building on this premise, the present study explores the potential of recommendation systems for policy matching across cities, illustrating how AI-driven personalization can be leveraged in urban and planning contexts.

## 3. Materials & Methods
### 3.1. Study Area and Materials

The study area includes U.S. cities with populations exceeding 100,000 that have adopted climate equity plans. The list of cities was compiled from Wikipedia, which reports estimated population figures for 2024. Among the 346 cities considered, 192 (55%) had regional climate equity plans. The corresponding documents were collected from publicly available repositories in PDF format. **Figure 1** illustrates the geographic distribution of cities with climate equity plans. While a small number of cities in Anchorage, Alaska, and Honolulu, Hawaii were identified, the majority of plans were concentrated in the contiguous U.S. For clarity of comparison and visualization, the analysis of policy practice differences focuses on the mainland U.S. Subsequently, these equity plans underwent a stepwise filtering process.

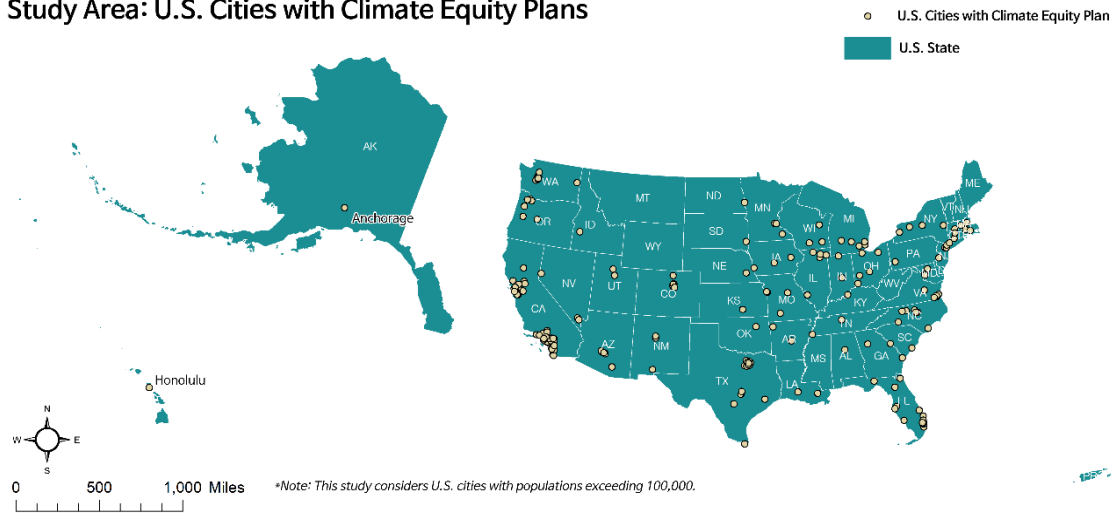

**Figure 1.** Study Area: U.S. Cities with Climate Equity Plans

The present study also collected job descriptions to identify the responsibilities of planners in professional practice. The posted jobs were web-scraped on October 23, 2025, from LinkedIn and Glassdoor. Job postings containing the keywords communication, community engagement, community outreach, crime, environment, equity, finance, housing, planning, policy, public administration, public health, transportation, and urban were collected separately. Using Natural Language Processing (NLP) techniques, irregular characters, markdown syntax, bullet points, URLs, punctuation, random letters, extra spaces, hyphens, and dashes were removed. In total, 69,116 U.S. job postings were collected. After reviewing all entries, only 199 job postings remained. There were instances of duplicate listings, including multiple openings for the same position and repeated postings across different job announcements. After removing these duplicates, a total of 83 job postings remained.

### 3.2. Methods

The research methodology of the present study consists of three stages. First, we conducted a text analysis of planning-related job titles and job descriptions. Word clouds were generated to visualize frequently occurring terms after removing stopwords (e.g., and, or, for, the, by, of, with, etc.). The cleaned job descriptions were transformed into term–document matrices using term frequency–inverse document frequency (TF-IDF) weighting. Latent Semantic Analysis (LSA), implemented via truncated Singular

Value Decomposition (SVD), was then applied to reduce dimensionality and extract latent topic structures. LSA-derived document–topic scores were used to examine thematic patterns across job descriptions, and the resulting topic intensities were visualized using a heatmap. The heatmap is used to visually examine the relative intensity and co-occurrence of LSA-derived topic scores across job descriptions, enabling comparison of thematic emphasis and identification of dominant and marginal roles in planning practice. For each topic, fifteen representative words were identified.

Additionally, the same text-analytic approach was applied to review climate equity plans. We first compiled all terms appearing in the tables of contents across the collected climate equity plans. Generic section headings such as introduction, results, and appendix were treated as stopwords, as they do not convey substantive thematic meaning. The refined set of terms was then analyzed using the same text-analysis procedure to identify themes that are commonly addressed in climate equity plans.

Second, a RAG approach was applied for the systematic review of climate equity policies. The RAG framework was adopted to improve factual accuracy and contextual completeness compared to direct large LLM prompting. It has been shown to mitigate hallucinations in LLMs and to improve the accuracy of factual information (Zhao et al., 2024). Policy documents in PDF format were loaded using the LangChain framework and segmented into text chunks to ensure comprehensive coverage of long documents. Texts were split using a recursive character-based strategy with a chunk size of 1,000 words and an overlap of 200 words to preserve semantic continuity across segments. Each text chunk was embedded using OpenAI's text-embedding-3-small model and indexed in a FAISS vector database. Document retrieval was conducted using a maximal marginal relevance strategy to balance relevance and diversity, with parameters set to $k = 5$, fetch_k = 20, and $\lambda = 0.7$. Here, k determines the number of retrieved text chunks used for generation, fetch_k defines the candidate retrieval pool, and $\lambda$ balances relevance and diversity, favoring relevance while avoiding redundant policy excerpts. In addition, random sampling was applied to further diversify retrieved contexts and reduce retrieval bias. For the generative component, the GPT-4o-mini model was employed with the temperature parameter fixed at 0 to ensure deterministic and reproducible outputs. The temperature parameter controls the variability of generated responses and is commonly interpreted as influencing creativity or novelty (Peeperkorn et al., 2024). Ki et al. (2025) set the temperature parameter to zero when measuring street walkability to ensure deterministic outputs. Similarly, Kornblith et al. (2025) used a zero temperature setting in clinical sentiment analysis to minimize randomness. Retrieved document chunks were then passed to ChatGPT using a structured prompt that constrained responses to the provided context and required explicit citation of source documents and page numbers.

The automated policy extraction process using ChatGPT was adapted from Deng et al. (2025), who conceptualized policy content as a hierarchical structure consisting of policies, strategies, and actions. **Figure 2** illustrates the workflow of policy extraction using an LLM-based RAG approach. Although document collection involved human oversight, we first applied an automated screening step to identify whether each document explicitly acknowledged challenges related to climate equity. ChatGPT was prompted to determine whether dedicated categories or sections addressing these challenges were present. When such sections were identified, ChatGPT was then instructed to extract relevant policies, strategies, and actions from the document. The extracted cases were compiled to identify prevailing contemporary practices. Our analysis focused on local practices related to transportation and energy, as these domains were most frequently addressed, as indicated by the table of contents analysis.[1] These policies generally aim to mitigate greenhouse gas emissions; for example, the transportation sector is the largest contributor to such emissions. This trend aligns with contemporary climate equity planning practices observed across diverse global contexts (Damsø et al., 2016; Deetjen et al., 2018; Wu, 2023).

Finally, the presence of each identified policy, strategy, and action was reassessed using structured binary prompts. This process was conducted across 20 predefined thematic categories. **Table 1 and 2** present the list of thematic categories for transportation- and energy-related policy practices identified in contemporary climate equity plans. As illustrated, actions are generally more contextual and descriptive in nature compared to policies or strategies. These themes are identified by ChatGPT based on a compiled database of policies, strategies, and actions.[2] To ensure credibility, ChatGPT was instructed to provide page numbers for all extracted statements and to respond with "I don't know" when uncertainty arose.[3] In total, the review process involved examining 21,489 pages comprising 5,834,348 words. The documents span the period from 2009 to 2025, with the majority of climate equity plans published between 2020 and 2025 (129 documents, representing 69%).

---

[1] From the table of contents, excluding stop words, the words that appear at least 2 times were 523. After excluding adjectives and general terms such as plan, goals, and strategies, we identified the top ten themes in order of frequency. The most common theme was climate (189 occurrences). This theme was followed by terms related to greenhouse gas emissions, including greenhouse gas emissions (211 occurrences), emissions (77), greenhouse (50), gas (47), and GHG (37). Other frequently appearing themes included energy (81), community (70), transportation (66), sustainability (62, consisting of 44 occurrences of sustainability and 18 of sustainable), resilience (33), water (30), land (30), engagement (27), and so on.

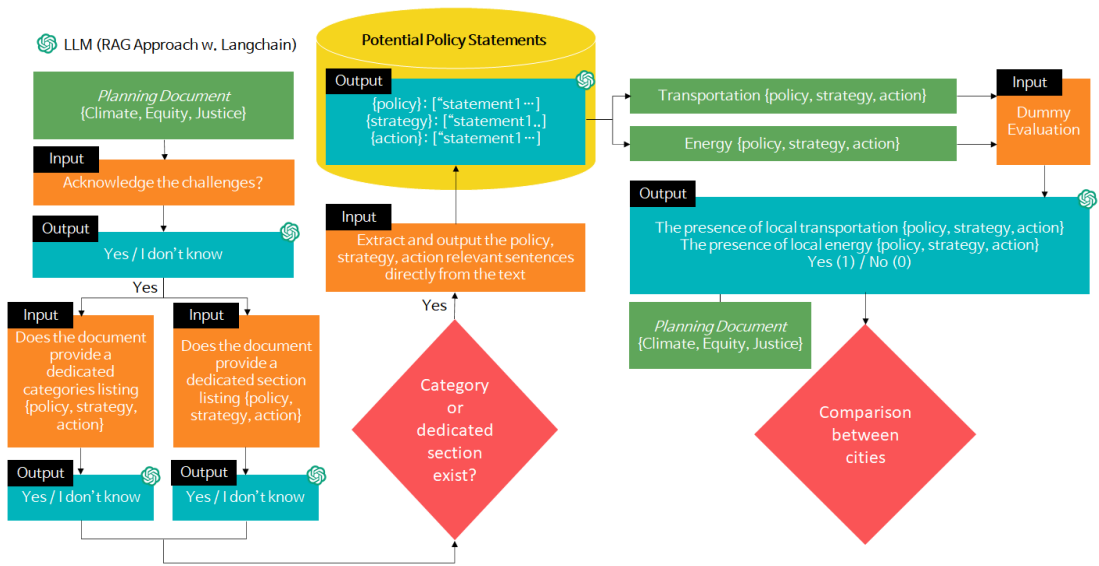

**Figure 2.** Flowchart of Policy Extraction and Evaluation Using an LLM-Based RAG Approach Built on LangChain

---

[2] While Deng et al. (2025) manually constructed a list of local policy determinants, our primary objective was to explore pathways toward automating the policy recommendation system.
[3] The identification of policies, strategies, and actions was based on semantic interpretation by an LLM rather than strict keyword detection. While this approach allows for greater flexibility in capturing substantively equivalent policy concepts expressed using different language, it may also introduce interpretive uncertainty. To mitigate this risk, the model was instructed to return page numbers and to respond with "I don't know" when sufficient evidence was not present.

**Table 1.** Identified Contemporary Themes in Transportation Policies, Strategies, and Actions from Climate Equity Plans

| Policy | Strategy | Action |
|---|---|---|
| Sustainable Transportation | Sustainable Transportation | Promote Bicycle Commuting |
| Transportation Equity | Transit System Improvement | Implement Carpool Programs |
| Multimodal Transportation Framework | Active Transportation | Expand Public Transit Services |
| Vehicle Emissions Standard | Low-Emission Vehicle | Reduce Vehicle Miles Traveled |
| Public Transit Accessibility | Transportation Demand Management | Encourage Active Transportation |
| Active Transportation | Bicycle Infrastructure Development | Develop Bicycle Lanes |
| Transportation Infrastructure Investment | Carpool and Vanpool Promotion | Enhance Transit Accessibility |
| Complete Streets Policy | Public Transit Expansion | Launch Transportation Awareness Campaigns |
| Bicycle and Pedestrian | Vehicle Idling Reduction | Install Electric Vehicle Chargers |
| Transportation Demand Management | Parking Management | Conduct Transportation Needs Assessments |
| Low-Emission Vehicle | Complete Streets Implementation | Implement Parking Management Solutions |
| Carpooling Incentive | Freight Efficiency | Support Vanpool Services |
| Transit-Oriented Development | Community Engagement in Transportation | Adopt Complete Streets Projects |
| Parking Management | Transportation Equity Strategies | Facilitate Community Workshops |
| Freight Transportation | Smart Mobility Strategies | Improve Transit Safety Measures |
| Transportation Safety | Integrated Transportation Planning | Encourage Use of Public Transit |
| Zero-Emission Vehicle | Green Transportation | Implement Traffic Calming Measures |
| Land Use and Transportation | Electric Vehicle Charging Infrastructure | Promote Shared Mobility Options |
| Transportation Resilience | Transportation System Optimization | Conduct Transportation Impact Studies |
| Smart Transportation | Land Use and Transportation Integration | Engage Stakeholders in Planning |

**Table 2.** Identified Contemporary Themes in Energy Policies, Strategies, and Actions from Climate Equity Plans

| Policy | Strategy | Action |
|---|---|---|
| Clean Energy | Renewable Energy | Implement Energy Efficiency |
| Renewable Energy Standards | Energy Efficiency | Promote Renewable Energy Projects |
| Energy Efficiency | Decarbonization | Conduct Energy Audits |
| Energy Equity | Electrification | Launch Clean Energy Initiatives |
| Decarbonization | Energy Conservation | Support Electrification of Buildings |
| Electrification | Grid Modernization | Develop Community Solar Projects |
| Energy Conservation | Distributed Energy | Enhance Energy Storage Solutions |
| Sustainable Energy | Clean Energy Transition | Facilitate Energy Education Workshops |
| Grid Modernization | Carbon Reduction | Adopt Energy Conservation Measures |
| Distributed Energy Resources | Energy Storage | Engage in Carbon Reduction Activities |
| Clean Energy Access | Community Energy | Install Renewable Energy Systems |
| Fossil Fuel Reduction | Sustainable Energy Development | Conduct Public Awareness Campaigns |
| Energy Storage | Energy Access | Implement Smart Grid Technologies |
| Carbon Pricing | Smart Grid | Support Energy Transition Efforts |
| Energy Transition | Energy Resilience | Promote Sustainable Building Practices |
| Net-Zero Energy | Energy Innovation | Conduct Energy Impact Assessments |
| Energy Resilience | Fossil Fuel Phase-Out | Encourage Energy Innovation |
| Community Solar | Energy Demand Management | Facilitate Stakeholder Engagement |
| Energy Education and Awareness | Green Building | Implement Demand Response Program |
| Green Building Energy | Public Engagement in Energy | Support Local Energy Initiatives |

Third, policy practice comparisons were conducted using a recommendation system framework. A content-based approach was employed, in which similarity was derived from the attributes of policy content itself rather than from observed interactions or diffusion patterns. Paired matrices representing cities and their associated policies, strategies, and actions were constructed. Cosine similarity was then used to identify the top five most similar city pairs. For each policy element, the rate of adoption among similar cities was calculated as the mean of binary indicators. Based on these similarity scores, we evaluated whether specific policies, strategies, and actions were commonly practiced or whether gaps existed across cities. Representative examples were presented to illustrate the results.

To facilitate interpretation and practical use of the recommendation results, we implemented an interactive search-based interface that allows users to query a target city and retrieve comparable peer cities along with recommended policy actions. Through this interface, users can input or select a city name and dynamically view the most similar cities identified by the cosine similarity metric, as well as policy, strategy, and action elements that are commonly adopted by peer cities but absent in the target city's plan. This search-based implementation enables transparent exploration of policy similarities and gaps across cities and supports reproducible inspection of the recommendation outcomes. The interface was designed as an analytical aid rather than a decision-making tool, providing an interpretable mechanism to examine how current climate equity policies vary across jurisdictions. Policymakers and public agencies would benefit from similar interfaces and tools to efficiently support policy narrative drafting. The interactive interface was implemented using ipywidgets within a virtual computing environment.

## 4. Results

### 4.1. Textual and Thematic Analysis of Planning and Policy Narratives

**Figure 3** illustrates the results of topic modeling based on text analysis of job descriptions and job titles extracted from job postings. The heatmap presents topic modeling scores derived from job descriptions, highlighting the relative prominence of each topic. Across job descriptions, the most frequently occurring terms were planning (876), experience (499), development (404), work (375), city (310), public (301), transportation (296), related (276), projects (242), information (219), and environmental (216), among others. Job titles predominantly included keywords directly related to planning roles, such as planner (54) and planning (29). Other commonly appearing terms included transportation (17), GIS (11), urban (11), director (10), assistant (9), analyst (9), environmental (7), city (5), and operations.

Based on the LSA-derived topic scores, Topic 1 was primarily centered on city (0.149), transportation (0.149), and development (0.141), reflecting general urban and transportation planning roles. Topic 2 was strongly associated with transportation (0.242), the Metropolitan Transportation Authority (MTA) (0.135), administration (0.105), and management (0.096), suggesting alignment with managerial or director-level positions (0.088). Topic 3 also emphasized transportation (0.220) but highlighted technical and regulatory expertise, including zoning (0.166), GIS (0.154), and software proficiency such as ArcGIS (0.124), along with knowledge of permits (0.119) and land use (0.104). Topic 4 was most strongly associated with intensive use of GIS (0.236) and a technical orientation (0.142), with an emphasis on data analysis (0.134). This topic also showed close associations with housing (0.136) and real estate (0.108). Topic 5 similarly reflected strong GIS usage (0.266), combined with expertise in transportation (0.210), environmental (0.156), regional (0.148), and housing (0.134) domains, indicating a multidisciplinary planning profile.

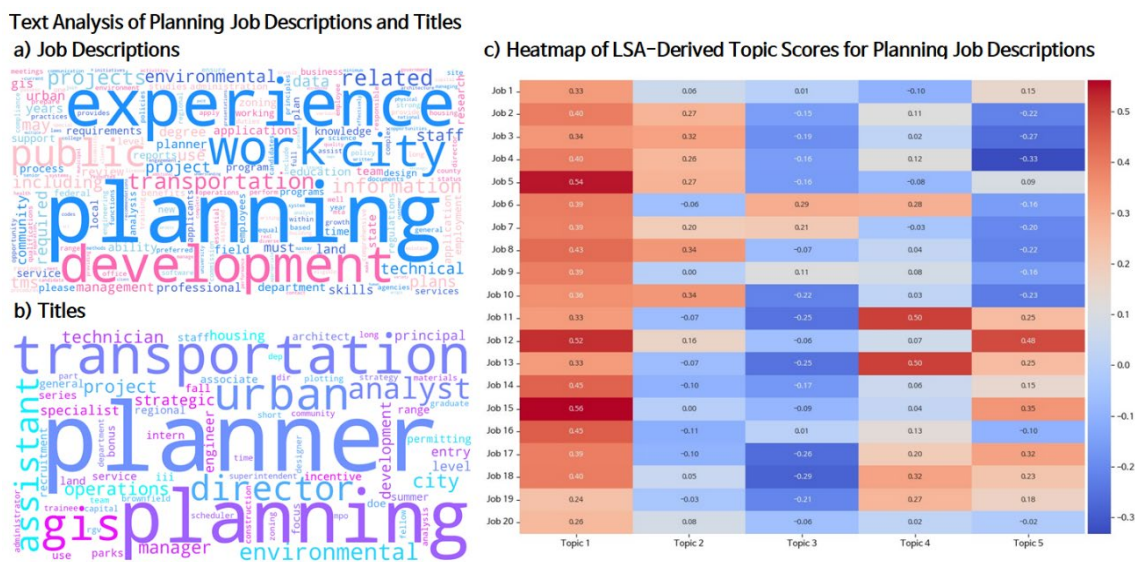

**Figure 3**. Text Analysis of Planning Job Descriptions and Titles: (a) Word Cloud of Job Descriptions; (b) Word Cloud of Job Titles; (c) Heatmap of LSA-derived Topic Scores across Planning Job Descriptions

**Figure 4** further illustrates the topic modeling and text analysis based on both the table of contents and the RAG-LLM–generated responses to the prompt: "What are the principal challenges and pressing issues addressed in this regional plan regarding the promotion of social equity and community advocacy?" Analysis of the planning documents directly extracted from the table of contents shows that climate (195), plan (118), and action (112) were the most frequently occurring keywords. Other prominent terms included energy (81), emissions (77), community (73), transportation (66), reduction (59), greenhouse (50), gas (47), waste (46), and sustainability (44). In comparison, the RAG-LLM–generated responses exhibited a stronger emphasis on equity-oriented language, with frequent occurrences of climate (171), communities (131), access (107), equitable (89), engagement (88), residents (76), income (67), health (65), change (59), vulnerable (58), action (58), populations (58), necessity (58), housing (55), transportation (53), low (52), environmental (51), strategies (50), enhance (50), and energy (50).

Each sentence in the generated responses was scored along the topic dimensions, and the top-ranked sentences for each topic were selected as representative statements, capturing the core semantic content of each topic in a concise and interpretable manner. Topics 1–5 represent distinct yet interconnected dimensions of climate equity discourse across the analyzed plans. Topic 1 reflects a broad climate action framing that emphasizes communities (0.171), access (0.138), equitable (0.120), and engagement (0.117). Topic 2 centers on food security, with food (0.192), access (0.119), and services (0.104) emerging as dominant terms. Topic 3 relates to the promotion of sustainability and long-term environmental considerations. Topic 4 is primarily associated with energy-related themes, showing the highest score for energy (0.313), while also incorporating systematic (0.113), historical (0.127), and inequities (0.123). Topic 5 is largely centered on food (0.500), with additional links to housing (0.139), health (0.112), and infrastructure (0.107).

**Table 3** presents representative phrases derived from topic scores, demonstrating that climate equity/justice-oriented plans span multiple planning domains, including environmental policy, food security, housing, transportation, energy, and infrastructure. The LLM successfully identifies underlying structural needs by explicitly acknowledging disadvantaged groups, such as low-income populations (e.g., LIDAC) and communities of color (e.g., BIPOC). Moreover, action-oriented policy items reflecting existing local conditions are consistently extracted, highlighting the applicability of the approach for diagnosing equity-focused planning priorities.

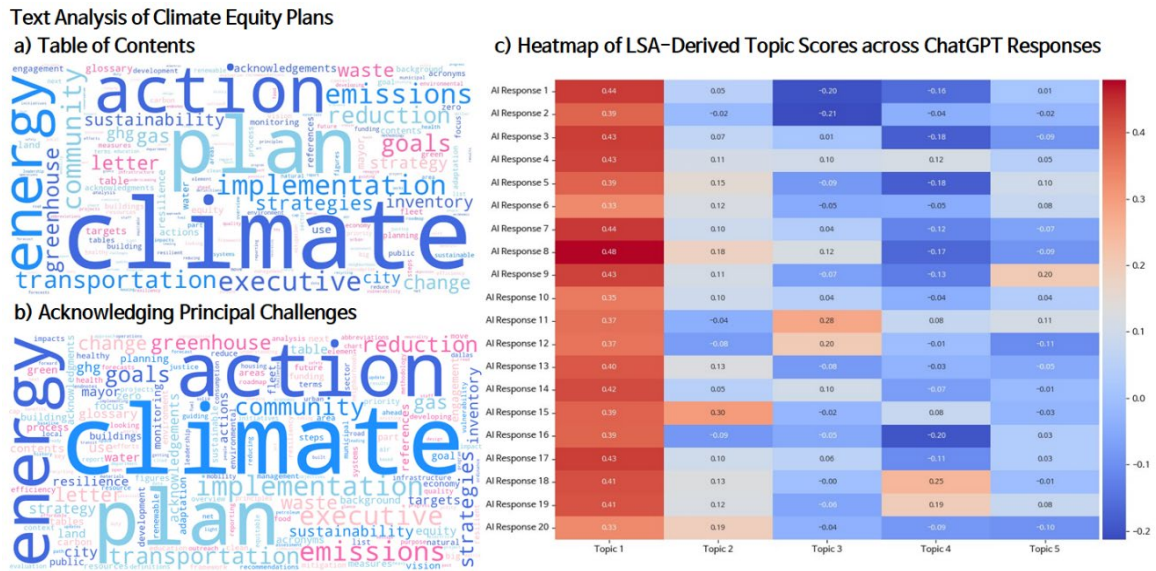

**Figure 4**. Text Analysis of Climate Equity Plans: (a) Table of Contents; (b) Acknowledging Principal Challenges; (c) Heatmap of LSA-derived Topic Scores across ChatGPT Responses

**Table 3**. Common Principal Challenges in Promoting Equity and Advocacy in Practice

| Topic | RAG-LLM Responses Generated Using LangChain |
|---|---|
| **Topic 1** Climate Access & Equity Promotion (keywords: climate, communities, access, importance, equitable, engagement, promotion, action) | *The principal challenges and pressing issues addressed in the regional plan regarding the promotion of social equity and community advocacy include the need for **direct engagement with LIDAC (Low-Income and Disadvantaged Communities) populations**, ensuring that climate action initiatives are inclusive and supportive of these communities.* |
| **Topic 2** Community Well-Being & Sustainable Services (keywords: communities, food, residents, access, sustainable, services) | *The principal challenges and pressing issues addressed in the regional plan regarding the promotion of social equity and community advocacy include the **disproportionate impacts** of climate change on marginalized and vulnerable populations, such as low-income communities and **BIPOC (Black, Indigenous, and People of Color) individuals**.* |
| **Topic 3** Climate | *The principal challenges and pressing issues addressed in the Westminster Sustainability Plan regarding the promotion of* |
| Sustainability and Policy Implementation (keywords: sustainability, climate, promotion, change, quality) | *social equity and community advocacy include the need for an **inclusive approach** that ensures diverse groups have equal rights, opportunities, and access to community capital programs and services.* |
| **Topic 4** Energy Transition and Environmental Justice (e.g., energy, food, climate, efficiency, dispariteis, inequities, emissions, healthy, systemic) | *The regional plan addresses several principal challenges and pressing issues related to social equity and community advocacy, including the need for enhanced community engagement in **climate action initiatives**, the promotion of **energy efficiency** and **renewable energy** access for all residents, and the **reduction of automobile emissions** through improved public transit and pedestrian infrastructure.* |
| **Topic 5** Food Systems, Housing, and Essential Services (e.g., food, climate, local, housing, production, health, action, infrastructure) | *The regional plan addresses several principal challenges and pressing issues related to social equity and community advocacy, including the need to overcome barriers to **food access** in underserved communities, such as **"food deserts,"** by encouraging the development of grocery stores and expanding food assistance benefits like SNAP/EBT/WIC (Supplemental Nutrition Assistance Program/Electronic Benefit Transfer/Special Supplemental Nutrition Program for Women, Infants, and Children).* |

**4.2. Semantic Review of Climate Equity Plans Using LLMs**

To further assess the linguistic characteristics of the extracted responses, a transformer-based language model was employed to derive semantic polarity scores. Specifically, a BERT-based classifier fine-tuned for binary sentiment classification was applied to the LLM-generated responses corresponding to policy sections, strategy sections, and action sections, as well as their respective categorizations. For each response, the model produced both a categorical label (positive or negative) and a confidence score. These outputs were transformed into signed polarity values to preserve directional information, allowing negative polarity to indicate statements emphasizing absence, ambiguity, or lack of explicit mention, and positive polarity to reflect affirming or articulated content. The semantic polarity analysis was not intended to evaluate policy effectiveness but to diagnose how explicitly or assertively climate or equity measures were articulated across planning documents.

In most cases, existing climate equity plans predominantly use affirmative language when listing policies, strategies, and action items. Among documents that did not include "I don't know" responses, negative classifications were identified in 19 policy-related phrases, 19 strategy-related phrases, and 17 action-related phrases at the section level. When aggregated by category, the corresponding counts were 17, 27, and 31, respectively.

**Tables 4** and **5** present representative samples classified as positive and negative cases with the highest confidence scores. These samples are derived from section-level phrases. The positive cases include relevant sentences from Albuquerque, Sacramento, and Cleveland, while the negative cases are drawn from Brownsville, South Bend, and Tampa. Affirmative cases typically include language that supports future directions and planned actions. In contrast, negative cases often reflect policy language that acknowledges limitations, challenges, or unresolved issues. For example, a phrase from Brownsville explicitly limits the potential impact of the policy. The case of South Bend similarly highlights explicit cost constraints. There are instances in which a dedicated section exists but lacks substantive detail, as illustrated by the case of Tampa, Florida.

**Table 4**. Representative Samples Classified as Positive with the Highest Confidence Scores

| Items | RAG-LLM Responses Generated Using LangChain |
|---|---|
| Policy | *from* Albuquerque, NM<br>• Yes, the document includes a dedicated section listing policies. The relevant sentences are: "Support local and state legislation that prioritizes urban infill, brownfield redevelopment and renovations, rather than new developments/new construction." and "Support consistent and timely adoption of local and state legislation that requires developers and home builders to continue to meet current energy standards for newly constructed or renovated buildings and homes." |
| Strategy | *from* Sacramento, CA<br>• Yes, the document provides a dedicated section listing strategies. It mentions several key City initiatives that address sustainability and equity, including the establishment of the City's Office of Diversity & Equity, the formation of the City Council's Racial Equity Committee, the establishment of the City's Office of Climate Action & Sustainability and the Climate Work Plan, and the development of an Inclusive Economic Development Strategy and Action Plan. |
| Action | *from* Cleveland, OH<br>• Yes, the document provides a dedicated section listing actions. It states that "Actions are clear, specific, actionable steps that the |

City, its partners, and its residents can take to complete objectives and put our community on the path to meeting our goals." Additionally, it mentions that "the City of Cleveland has analyzed each action using the Evaluative Framework laid out in the previous chapter" and identifies "a handful of Priority Actions within each Focus Area that the City will focus on implementing immediately." Furthermore, it highlights a set of Cross-Cutting Priorities essential for advancing climate justice, which include Circular Economy, Empowerment, Equity, Good, Green Jobs, and Resilience.

Table 5. Representative Samples Classified as Negative with the Highest Confidence Scores

| Items | RAG-LLM Responses Generated Using LangChain |
|---|---|
| Policy | *from* Brownsville, TX<br>• Yes, the document includes a dedicated section listing policies. The relevant sentences are: "No public money incentives will be paid to companies that do not pay living wages. Living wage clause will be included in municipal purchase contracts. All contracts awarded by the municipality (or by public entities in Brownsville) would require contractors and all sub-contractors to pay living wages to employees. For new business, bidding credits will be awarded to bidders that pay living wages. A bidder that pays living wages would get a credit in their submitted bid equal to a specified percentage of the submitted bid." This is found in the section titled "Strategic Initiatives" on page 7. |
| Strategy | *from* South Bend, IN<br>• Yes, the document includes a section on strategies. The relevant sentences are: "While useful climate strategies can be pulled from around the globe, this plan prioritizes solutions that most clearly fit South Bend's institutions, demographics, natural environment, regional economy, and infrastructure. Oftentimes, the highest-impact strategies are not the most feasible. Considering the importance that this plan be actionable for the city, strategies were prioritized based on how achievable they were for the community, municipality, and other agencies, and whether they proved to be too expensive or at the cost of other priorities, like quality of life, access, and economic opportunity." This information is found in the section titled "INVENTORY AND METHODOLOGY" on page 6. |
| Action | *from* Tampa, FL<br>• Yes, the document includes a dedicated section listing actions under the "List of Projects, Programs, Policy and Administrative Changes" on page 146. However, the specific actions are not detailed in the provided context. |

## 4.3. Identification of Similarities and Gaps in Climate Equity Plans and Practices

In the evaluation of transportation and energy policy, strategy, and action themes within climate equity plans, no cases achieved a full score of 20 out of 20 for the respective criteria. On average, transportation themes scored 7 for policy, 6 for strategy, and 8 for action across the evaluated cities. Land use and transportation integration emerged as the most common transportation policy theme, with transportation resilience receiving the least attention. For transportation strategies, electric vehicle charging infrastructure was the most common theme, whereas smart mobility and integrated transportation planning were the least represented. For transportation actions, the installation of additional electric vehicle charging infrastructure was most frequently identified, while launching transportation awareness campaigns, conducting needs assessments, and conducting impact studies were the least commonly observed actions.

Across the energy themes, the average scores were 9 for policy, 8 for strategy, and 6 for action across the evaluated cities. Energy efficiency was the most frequently addressed theme in energy policy, while grid modernization was the least addressed. Renewable energy was the most common focus for energy strategies, while energy innovation and grid modernization received the least attention. Among action items, implementing energy efficiency measures was most common; however, conducting formal energy impact assessments was observed in only one case.

Based on the evaluation results, we constructed a content-based recommendation system. Recommendations are generated based on the similarity of evaluation patterns across action-related indicators. It is important to note that this similarity does not reflect geographic proximity between cities; rather, it indicates similarity in the action patterns recorded in the evaluation data. Because action items are more detailed and context-specific, we demonstrate a matching exercise in which cities with high scores are compared to others with similar practices. **Figure 6** illustrates the retrieval of transportation action items for the case of Las Vegas. Las Vegas exhibited analogous transportation action patterns to those of Chico, Ann Arbor, Berkeley, Richmond, and Long Beach. Across these cities, common practices include promoting shared mobility options, facilitating community workshops, and launching transportation awareness campaigns. However, the current documentation appears to lack action items explicitly focused on enhancing transit accessibility, supporting vanpool services, and conducting needs assessments. Similarly, **Figure 7** demonstrates the search for comparable cases for energy-related action items using Fort Lauderdale as the reference city. Fort Lauderdale shares similarities in practice with Saint Paul, Des Moines, Sterling Heights, Columbia, and Buffalo. Commonly observed actions include supporting energy transition efforts and facilitating stakeholder engagement. In contrast, actions such as launching clean energy initiatives, offering energy education workshops, adopting energy conservation measures, and conducting public awareness campaigns are likely missing.

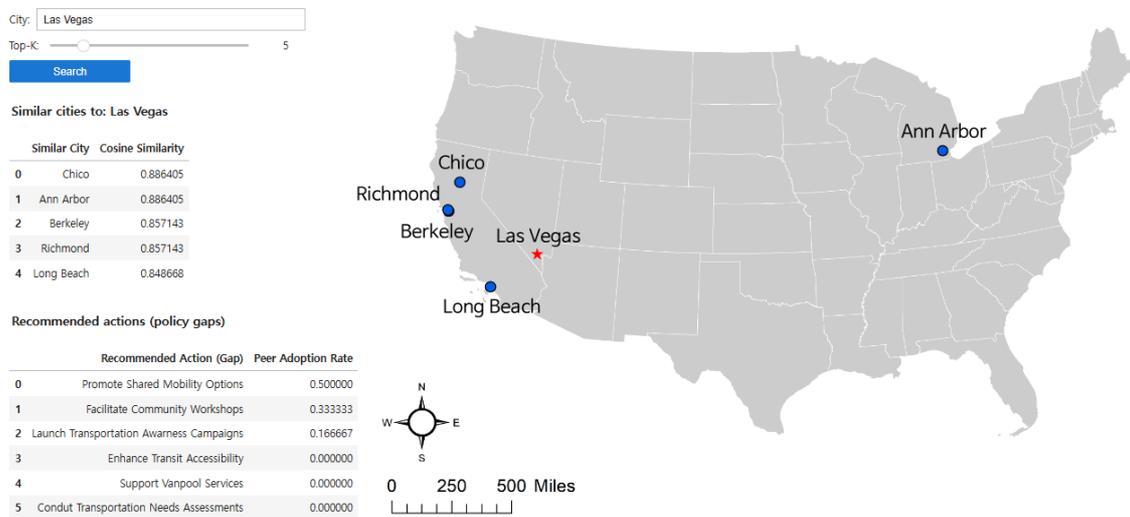

**Figure 6**. Demonstration of a Recommendation Search Engine for Transportation Actions

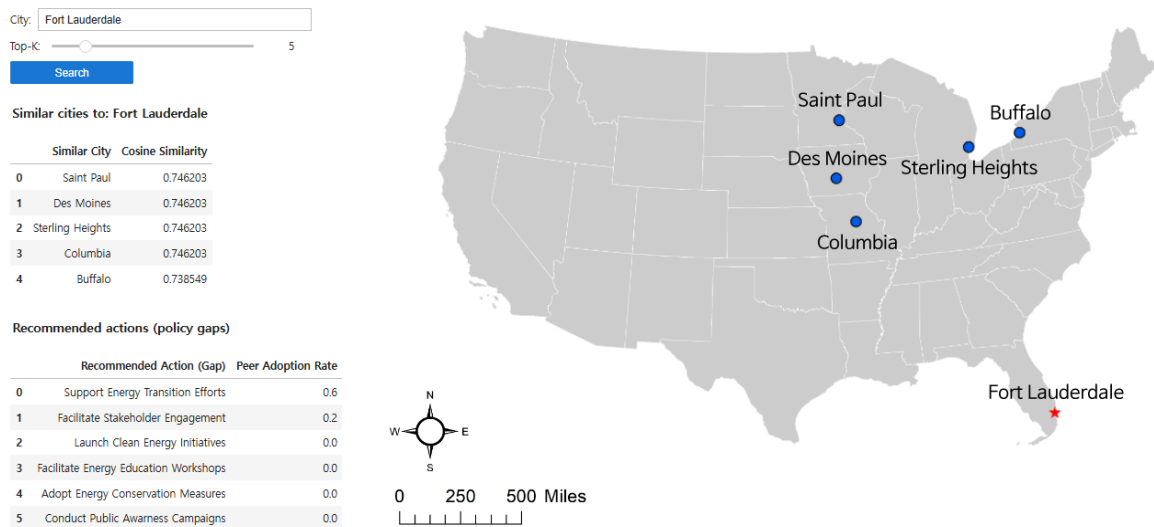

**Figure 7**. Demonstration of a Recommendation Search Engine for Energy Actions

## 5. Discussion

This study demonstrates how AI-based text analysis and recommendation systems can support the comparison, diagnosis, and learning of planning practices for climate equity. While AI does not replace planners' normative judgment, the findings show that such systems can function as analytical tools that reveal patterns embedded in large volumes of policy documents and expand the range of policy options available to practitioners. The analytical findings are as follows.

First, in the era of AI, this study revisits the evolving role of planners and their professional responsibilities. Explicit references to AI were largely absent from the analyzed job postings, despite frequent requirements for GIS and data analysis skills. Domain-specific planning areas such as transportation, environmental planning, land use, and housing remained central across positions. These roles were not confined to traditional planning job categories but extended across sectors, including business, management, finance, education, social services, technology, intelligent transportation systems, and logistics, and were observed in both public and private organizations. Taken together, the topic modeling results suggest that planning jobs are evolving toward a hybrid professional model that combines coordination-oriented generalist roles with specialized managerial and technical functions. Transportation remains a core domain; however, its professional expression is increasingly differentiated. At the same time, GIS and data analytics have become structural competencies rather than auxiliary skills. Collectively, these patterns indicate that contemporary planning practice increasingly requires the integration of normative judgment with technical and regulatory expertise.

Despite rapid advances in GenAI, planning jobs continue to emphasize human-centered communication and judgment as core professional responsibilities. An examination of job descriptions shows that planners are commonly expected to perform communicative roles, including managing and evaluating strategies, supporting departmental functions, and distributing information while anticipating and correcting misinformation. Such responsibilities align with Forester's (1982) view of planners as communicative actors, underscoring the enduring need for human-to-human interaction in planning practice. While recent advances in GenAI and LLMs are reshaping plan-making processes and affecting workforce composition (Zheng et al., 2025), positions that require human oversight, interpretation, and trust-building are likely to persist.

Similar to participatory planning traditions, emerging approaches to participatory algorithm design point toward new forms of algorithmic governance (Lee et al., 2019). The adoption of these tools depends on varying levels of trust, reliance, and aversion toward AI systems (Klingbeil et al., 2024). Consequently, domain-specific planning roles are expected to continue, with emerging responsibilities for the responsible use and governance of AI becoming an integrated and defining component of planning practice. However, these responsibilities are likely to be carried out more effectively by more experienced professionals, which may raise barriers to entry for junior practitioners in the labor market (Brynjolfsson et al., 2025).

Second, this study identified common elements in climate equity plans across the U.S. and applied semantic analysis to examine their content. Topic modeling results based on tables of contents confirm that transportation and energy, particularly in relation to greenhouse gas emissions reduction, are consistently emphasized policy domains. These findings demonstrate the translation of established theoretical frameworks into planning practice (Damsø et al., 2016; Deetjen et al., 2018; Wu, 2023). Based on the semantic review and sentiment classification of direct quotations from the documents, policy texts predominantly employ affirmative language, with relatively few instances of negative semantics used to acknowledge limitations or challenges.

In practice, electric vehicle charging emerges as a recurring area of attention within transportation-related policy content. Despite political pushback against electric vehicles, their adoption continues to grow; however, this expansion also raises equity concerns related to disproportionate access to charging infrastructure and electric vehicle ownership (Choi et al., 2024). Addressing these challenges is integral, and policymakers should consider leveraging overlooked action items such as public awareness campaigns, needs assessments, and impact studies. For energy-related action items, renewable energy is widely acknowledged as a priority; however, technical dimensions, such as grid modernization and efficiency improvements, are less frequently addressed, possibly due to the higher levels of specialized knowledge required in electrical engineering and grid systems. Based on this trend, ensuring that renewable energy systems are environmentally sustainable across their full life cycle is critical (Pehnt, 2006). Moreover, technical initiatives such as grid modernization encompass multiple interrelated elements, including infrastructure, operational processes, business models, and regulatory actions (Agüero et al., 2017). Addressing these challenges effectively requires engagement from multiple stakeholders to design more concrete and actionable policy interventions.

We further demonstrated that the use of LLMs fine-tuned with RAG frameworks can effectively support identification of community advocacy in equity planning. By implementing RAG through LangChain and applying deterministic parameter settings, the approach improves accuracy and reduces hallucination risks commonly associated with direct LLM use (Zhao et al., 2024). This capability is particularly important in the normative context of equity planning, where explicitly recognizing disadvantaged groups and their spatial characteristics is essential for meaningful implementation (Brooks, 2002; Loh and Kim, 2021). When policy documents are embedded in a vector database (e.g., using LlamaIndex), the system enables planners to retrieve localized, context-specific information based on targeted queries, further mitigating hallucination risks compared to standalone LLM services. Similar to how human planners develop plan equity evaluation tools through systematic content analysis (Loh and Kim, 2021), this approach can be extended to perform comprehensive evaluations at significantly greater speed and scale.[4] The use of affirmative versus negative language by virtual planners can be readily adjusted through the tuning of temperature parameters. However, prior to institutional deployment, it remains necessary to assess levels of trust in these systems through validation of accuracy (Deng et al., 2025) and the application of risk metrics tailored to AI governance contexts (Mohawesh et al., 2025).

Third, this study demonstrates the use of LLMs to develop a recommendation system for matching policy content across climate equity plans in different cities. The system allows users to adjust the number of recommended cities based on similarity, and it can be extended to evaluate policies, strategies, and action items separately. The results suggest that policy practices often exhibit geographically patterned similarities. For example, when examining similarities in transportation-related action items for Las Vegas, several cities in California, including Chico, Richmond, Berkeley, and Long Beach, were identified as comparable cases. Similarly, when Fort Lauderdale was used as a reference city for energy-related action items, a cluster of cities located primarily in the eastern U.S. emerged, including Saint Paul, Des Moines, Columbia, Sterling

---

[4] This approach requires supporting technical infrastructure, including server resources, vector databases, and data management systems for institutional deployment.

Heights, and Buffalo. The observed geographic clustering does not imply causality but highlights how regional contexts may influence the convergence of policy practices across jurisdictions. This pattern is consistent with the view that policymaking is mobile and shaped through policy conversations that circulate, with variation, across geographic contexts (Temenos and McCann, 2013).

The recommendation system developed in this study offers practical value as a policy learning and diagnostic tool. Rather than prescribing optimal policies, it enables planners to identify peer cities with similar policy profiles and examine which actions are present or absent in their plans. By surfacing overlooked or underutilized action items from comparable jurisdictions, the system can support reflective planning, peer benchmarking, and iterative policy refinement. The system is meant to help planners make decisions, not take their place. It provides an evidence-based starting point for exploring alternative policy actions while allowing planners to account for local context, institutional capacity, and political feasibility. While large-scale urban systems in countries such as the U.S. and China may benefit from such recommendation systems for policy management, cities operating at smaller scales may derive comparable, if not greater, value. For example, countries experiencing significant declines in birth rates, including South Korea, as well as shrinking cities and suburban areas in parts of Japan and China, could leverage these systems to support compact city planning and governance.

## 6. Conclusion

Questions about how AI can assist policymaking gave rise to this study. We envision urban systems evolving toward greater intelligence, with existing tools and infrastructure increasingly integrated with LLMs. Private companies have already begun adopting such technologies in their operations, which has intensified the need for orchestrating multiple AI agents and supporting infrastructure to ensure reliable performance, including next-generation communication networks such as 6G (Tarkoma et al., 2023). Planners primarily operate within the public sector and carry obligations to promote public goods (Brooks, 2002). In this context, we envision planning agencies slowly adapting to these technological shifts. Specifically, this study focuses on climate equity plans across the U.S. and employs ChatGPT to conduct cross-city comparisons and develop a policy recommendation system. In parallel, we examine planning-related job postings to reassess the evolving roles of planners in the era of AI prior to conceptualizing the notion of virtual planners.

This study contributes to the field by demonstrating how LLM-based RAG systems can support comparative learning, diagnostic evaluation, and policy recommendation in climate equity planning. By structuring policy documents into policy–strategy–action components and implementing a content-based recommendation system, the study moves beyond keyword-based assessments toward actionable policy insights. Importantly, the findings position AI not as a substitute for planners' normative judgment but as an infrastructure that augments cross-city learning while highlighting institutional, workforce, and governance considerations for public-sector adoption. We further suggest directions for urban AI planners regarding the institutional adoption of AI-embedded systems in practice.

Nevertheless, several limitations should be noted. First, responses generated using the RAG approach may vary depending on text chunking strategies, including chunk size. Despite mitigating this issue by setting the temperature parameter to zero to ensure deterministic outputs, some variability may persist. Second, multiple LLMs and different versions of ChatGPT exist, and model updates may influence results.[5] Third, while alternative frameworks such as LlamaIndex can also be used to construct RAG pipelines, their effectiveness for policy review was not evaluated in this study.[6] Fourth, policy identification by ChatGPT relies on semantic interpretation, which may differ from classifications produced through human auditing. Fifth, the demonstrated recommendation system does not incorporate geographic information; although this study adopts a content-based approach, performance may differ if collaborative filtering or spatially informed methods are applied. These limitations highlight important directions for future research.

---

[5] We adapted the framework proposed by Deng et al. (2025), who demonstrated that a ChatGPT-based policy extraction approach achieved an average recall of 0.795 while reducing human document review workload by 44.0% to 99.0% across planning documents. We do not claim equivalence to human auditing.

[6] Implementing vector-based RAG pipelines typically requires substantial storage and memory resources to maintain document embeddings, which may pose practical constraints depending on institutional computing capacity.

**Funding Details**

The authors declare that this research received no external funding.

**Acknowledgements**

The authors declare no conflicts of interests.